\documentclass[aps,pre,twocolumn,floatfix,showpacs]{revtex4}
\usepackage{graphicx,epsfig}
\usepackage{bm}
\usepackage{amsmath, amsfonts, amssymb}
\bibstyle{apsrev.bst}

\newcommand{\etal}{\textit{et al.}}

\newcommand{\fig}{Figure}

\newcommand{\bc}{\begin{center}}
\newcommand{\ec}{\end{center}}
\newcommand{\be}{\begin{equation}}
\newcommand{\ee}{\end{equation}}
\newcommand{\ba}{\begin{array}}
\newcommand{\ea}{\end{array}}
\newcommand{\beqn}{\begin{eqnarray}}
\newcommand{\eeqn}{\end{eqnarray}}
\newcommand{\degree}{\ensuremath{^{\circ}}}
\newcommand{\Ginit}{\ensuremath{G^{\text{init}}}}

\begin{document}
\title{Effective affinities in microarray data}
\author{T. Heim}
\affiliation{Interdisciplinary Research Institute c/o IEMN, Cit\'e
Scientifique BP 60069, F-59652 Villeneuve d'Ascq, France}
\author{J. Klein Wolterink}
\affiliation{Institute for Theoretical Physics, University of Utrecht,
Leuvenlaan 4, 3584 CE Utrecht}
\author{E. Carlon}
\affiliation{Interdisciplinary Research Institute c/o IEMN, Cit\'e
Scientifique BP 60069, F-59652 Villeneuve d'Ascq, France}
\affiliation{Ecole Polytechnique Universitaire de Lille, Cit\'e
Scientifique, F-59655 Villeneuve d'Ascq, France}
\author{G.~T.~Barkema}
\affiliation{Institute for Theoretical Physics, University of Utrecht,
Leuvenlaan 4, 3584 CE Utrecht }

\date{\today}

\begin{abstract}
In the past couple of years several studies have shown that hybridization
in Affymetrix DNA microarrays can be rather well understood on the
basis of simple models of physical chemistry. In the majority of the
cases a Langmuir isotherm was used to fit experimental data. Although
there is a general consensus about this approach, some discrepancies
between different studies are evident. For instance, some authors have
fitted the hybridization affinities from the microarray fluorescent
intensities, while others used affinities obtained from melting
experiments in solution. The former approach yields fitted affinities
that at first sight are only partially consistent with solution values.
In this paper we show that this discrepancy exists only superficially:
a sufficiently complete model provides effective affinities which are
fully consistent with those fitted to experimental data.  This link
provides new insight on the relevant processes underlying the functioning
of DNA microarrays.
\end{abstract}


\maketitle

\section{Introduction}

In all living cells the genes are transcribed, i.e., copied into messenger
RNA (mRNA), at different rates \cite{albe02_sh}. These rates depend on the
type of cell, on the stage of the cell life cycle and on other external
stimuli, like changes of pH, temperature or on the presence of chemicals.
The abundance of a specific mRNA defines the so-called gene expression
level.  It is of central importance to understand when, in which tissue
and in which amount a given gene is expressed.  This knowledge is for
instance crucial in understanding several diseases that originate from
deregulations in the gene transcription process, i.e., those pathologies
triggered by genes which are overexpressed or underexpressed.

DNA microarrays have become pivotal devices in molecular biology as they
allow a genome-wide screening of gene expression levels in a single
experiment. Both commercial and home made microarrays are nowadays
available. One of the leading companies in the DNA-microarray market
is Affymetrix, which produces high-density oligonucleotide microarrays
\cite{lips99_sh}. In Affymetrix arrays, photolitographic techniques are
used to grow on a solid substrate single-stranded DNA sequences which
are 25 nucleotides long; these are normally referred to as \emph{probes}.
The array is placed in contact with a solution containing RNA molecules,
i.e., the \emph{targets}, extracted from biological samples. Those
targets that are complementary to probe sequences tend to bind to
these, a process known as \emph{hybridization}.  Biotin molecules are
attached to a fraction of the nucleotides in the target sequences. Once
hybridization has occurred and the unbound targets are washed away,
streptavidin molecules, which carry fluorescent labels, are added
to the solution. The latter bind with high affinity to the biotin so
that the amount of hybridized probe-target duplexes can be determined
experimentally by optical measurements.

Two specific aspects of Affymetrix arrays are: 1)~Several probes are
complementary to the same target molecule (these probes form the so-called
probe set) and 2)~Each perfect matching (PM) probe has a partner probe
which differs by a single nucleotide in the middle position, the so-called
mismatch (MM) probe.  The use of multiple probes for the same target RNA
increases the reliability of the determination of gene expression levels
in Affymetrix arrays, which are obtained from simultaneous measurements of
several fluorescent signals. The signals measured from MM probes can be
used as test for the quality of the hybridization experiment.  Usually,
one expects that PM probes give a stronger signal than the corresponding
MM probes. However, ``bright mismatches", i.e., higher signals from MM
than PM probes, are observed quite frequently \cite{naef03}.

The hybridization of complementary strands in solution, or the reverse
process of DNA/RNA melting, has been widely investigated in the past
years \cite{bloo00}. Measurements of melting temperatures of short
oligonucleotides have yielded estimates
of the enthalpy and entropy differences $\Delta H$ and $\Delta S$
between a double helix and the two separate strands.  It turns out that
$\Delta H$ and $\Delta S$ can be well approximated by a sum over local
terms depending on pairs of neighboring nucleotides, plus eventual
boundary terms. This defines the so-called nearest-neighbor model
\cite{bloo00}. Table \ref{table_DG} gives an example of nearest-neighbor
free energy parameters obtained from measurements of melting temperatures
of DNA/RNA duplexes in solution.  The free energy differences are obtained
from $\Delta G = \Delta H - T \Delta S$, assuming that the experimentally
measured $\Delta H$ and $\Delta S$ are temperature independent.

\begin{table}[t]
\caption{The stacking free energy parameters $\Delta G$ for
RNA/DNA hybrids measured in solution at a salt concentration of $1$ M NaCl and
at 45\degree \cite{sugi95_sh}. The upper strand
is RNA (with orientation 5'-3') and lower strand DNA (orientation 3'-5').
The helix initiation energy is $\Delta \Ginit = 3.14$ kcal/mole.}
\begin{ruledtabular}
\begin{tabular}{cc|cc}
Sequence     & $-\Delta G$ (kcal/mole)  & Sequence     & $-\Delta G$ (kcal/mole)  \\
\hline
&&&\\
${\rm rAA} \atop {\rm dTT}$ & 0.83 &
${\rm rAC} \atop {\rm dTG}$ & 1.99 \\
&&&\\
${\rm rAG} \atop {\rm dTC}$ & 1.62&
${\rm rAU} \atop {\rm dTA}$ & 0.70\\
&&&\\
${\rm rCA} \atop {\rm dGT}$ & 0.70&
${\rm rCC} \atop {\rm dGG}$ & 1.92\\
&&&\\
${\rm rCG} \atop {\rm dGC}$ & 1.32&
${\rm rCU} \atop {\rm dGA}$ & 0.73\\
&&&\\
${\rm rGA} \atop {\rm dCT}$ & 1.21&
${\rm rGC} \atop {\rm dCG}$ & 2.56\\
&&&\\
${\rm rGG} \atop {\rm dCC}$ & 2.65&
${\rm rGU} \atop {\rm dCA}$ & 0.93\\
&&&\\
${\rm rUA} \atop {\rm dAT}$ & 0.42&
${\rm rUC} \atop {\rm dAG}$ & 1.31\\
&&&\\
${\rm rUG} \atop {\rm dAC}$ & 1.37&	
${\rm rUU} \atop {\rm dAA}$ &-0.08\\
&&&
\end{tabular}
\end{ruledtabular}
\label{table_DG}
\end{table}

The hybridization process in microarrays is not identical to that in
solution, as one of the two strands is surface-bound. A review of recent
work on the hybridization on surface immobilized DNA \cite{levi05}
shows that the rate constants for hybridization are lower than those
predicted by the nearest neighbor model in solution. The comparison
was done with experiments with a single species target and
probes of equal length \cite{okah98_sh,nels01_sh,pete02}.

Several studies \cite{naef03,vain02,deut04,held03,halp04,haga04,burd04}
recently discussed the role of the Langmuir isotherm and variants
thereof in connection with DNA microarrays.  Research toward a
physics-based modeling of hybridization in Affymetrix arrays can
roughly be divided into two approaches.  The first approach is to
identify empirical functions with many degrees of freedom, that
are fitted to experimental data~\cite{naef03,bind05}. The other
approach is molecular-based and employs the hybridization energies in
solution; it then requires a rescaling of parameters like the effective
temperature~\cite{held03,carl06}.  The aim of this paper is to link
these two apparently different viewpoints.  We shall show indeed that,
when the appropriate quantities are compared, i.e. the {\it effective
affinities}, the two models yield fully consistent results.

This paper is organized as follows: Sec.~\ref{sec:affyexpdata}
reanalyzes the binding affinities as introduced by Naef and
Magnasco~\cite{naef03} and Binder and Preibisch~\cite{bind05}. We
carry out a sensitivity analysis and show which features are robust
and which are sensitive. In Sec.~\ref{sec:affymodeldata}, effective
affinities are calculated using a molecular model based on the binding
free energies of Sugimoto \etal\cite{sugi95_sh} and the extension by
Carlon and Heim~\cite{carl06}.  From this model, the influence of
different additions to the molecular model on the effective affinities
is calculated and analyzed.
Section \ref{sec:conclusion} concludes the paper and summarizes the main
results.

\begin{figure}[t]
\includegraphics[width=5.8cm]{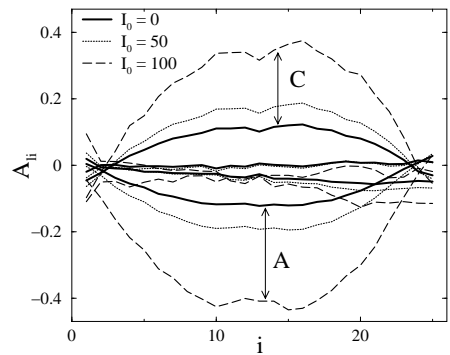}
\caption{Position-dependent effective affinities fitted from Affymetrix
data of the HGU95A chipset. Three different background values are 
subtracted: $I_0 = 0$, $50$ and $100$. The three topmost curves are
the affinities for nucleotides C and the three lowest curves for
the nucleotides A. The affinities for T and G are almost degenerate.}
\label{FIG01}
\end{figure}

\section{Effective affinities for Affymetrix arrays}
\label{sec:affyexpdata}

We turn now to the determination of the effective affinities from the
analysis of Affymetrix data. We follow here and further the procedure
originally introduced by Naef and Magnasco~\cite{naef03} and extended
more recently by Binder and Preibisch \cite{bind05}.

Naef and Magnasco fit the brightness $B$ 
of perfect-matching probes as a function of their sequence composition:
\begin{equation}
\ln \left( \frac{B}{\mbox{[RNA]}} \right) = \sum_{li} S_{li} A_{li},
\label{eq:naef}
\end{equation}
where $l=A,C,G,T$ is the letter index and $i=1,\dots 25$ the position
along the probe. $S_{li}$ is a boolean variable equal to 1 if
the probe sequence has letter $l$ at position $i$ and 0 otherwise, and
thus $A_{li}$ are per-site, per-letter affinities. The median of the PM
brightnesses [RNA] is used in this expression as a surrogate for the
RNA concentration, which is not known for most Affymetrix data. 

In Affymetrix experiments, the brightness $B$ will saturate, once
the majority of the probes are bound to targets. Capturing such
saturation requires the use of Langmuir isotherms; the approach above
(eq.~(\ref{eq:naef})) neglects saturation effects, and hence is only
expected to work in the so-called Henry regime~\cite{fics1} signified
by brightnesses much lower than the maximal value. Since only few probes
reach saturation, neglecting saturation is justifiable.

The experimentally measured fluorescence intensity $I_s$ of a probe with
sequence $s$ does not approach zero at zero concentration of the matching
target: there is a background signal, probably due to non-specific binding.
To take this into consideration, we distinguish two contributions to the
fluorescence intensity: a constant background intensity $I_0$ and the
brightness $B$ due to specific binding:
\begin{equation}
I_s=I_0+B,
\label{eq:naef_bg}
\end{equation}
in which $B$ is the brightness as in eq.~({\ref{eq:naef}).  We tried
different background subtractions schemes in order to test the
robustness of the data. Fig.~\ref{FIG01} shows the position-dependent
affinities $A_{li}$ obtained from fitting the experimental data to
eqs.~(\ref{eq:naef}) and (\ref{eq:naef_bg}) for background intensities
of $I_0 = 0$, $50$ and $100$ (constant background level). In the fit,
the distance of the data to the model was minimized in the logarithmic
scale.  We note that although the shape of the fitted position-dependent
affinities remain the same in the three cases, the amplitudes vary by a
factor of $4$. In all cases the shape is consistent with what was found
in Refs.\ \cite{naef03,bind05}: the position-dependent affinities are
approximately symmetrical with respect to the central position of the
probe ($i=13$) and the highest affinity is for nucleotides C and the
lowest for A in the probe sequence.  The affinities for the G and T
bases are almost degenerate and show less position dependence than the
affinities for the C and A bases.

In the case of $I_0=0$ we have a rather low signal. This is somehow
expected as in that case the non-specific part of the signal may dominate,
which induces a loss of specificity. 
When higher values of $I_0$ are taken, a non-trivial signal
starts to emerge. As $I_0$ increases, the amplitude of the strongest
effective affinity increases to $0.2$ and $0.4$ for respectively $I_0=50$
and $100$.

\begin{figure}[t]
\includegraphics[width=5.8cm]{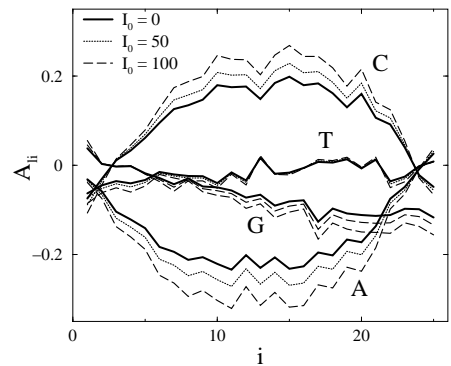}
\caption{As in Fig.~\ref{FIG01}, but disregarding all the data for probe
sets with an average intensity below $I=500$. The effective affinities
are less sensitive to the choice of $I_0$, compared with the fits of
Fig.~\ref{FIG01}.}
\label{FIG02}
\end{figure}

In Fig.\ \ref{FIG02} we plot the fitted affinities $A_{li}$ for
probe sets with an average intensity above $500$. This case corresponds
to signals well above the background level and thus the results should
be weakly dependent of the value of $I_0$ chosen, as is indeed the
case. 

As mentioned above, using the median of the PM brightnesses [RNA]
as an estimate for the RNA concentration is the only thing one can do
in the absence of knowledge of its true value.  Affymetrix, however,
performed a set of experiments in which some target sequences are added
in solution (spiked-in) at a known concentration. The results, known as
the Latin square data set, are publicly available from the Affymetrix web
site \cite{affy}.  We used these data to refit the effective affinities
from eq.~(\ref{eq:naef}), using the true target concentration $c_s$
of sequence $s$, rather than the median of the intensities.  Due to the
large number of parameters, this procedure yields typically values of
$A_{li}$ that are too noisy.  To limit the number of fitting parameters we
therefore have fitted $A_{li}$ only at some fixed positions $i=1,4,7,10
\ldots 25$ and taken for the other values of $i$ a linear interpolation
between the two fitted numbers. Note that the Latin square set also
contains a series of reference intensities measured in absence of
the transcripts in solution (i.e.\ $c_s=0$), a
procedure that yields a direct estimate of the background signal $I_0$.
The position-dependent affinities obtained from the fitting of the Latin
square set are shown in Fig. \ref{FIG03}. The results,
although still somewhat noisy, follow the general trend already shown
in Figs. \ref{FIG01} and \ref{FIG02}.

\begin{figure}[t]
\includegraphics[width=5.8cm]{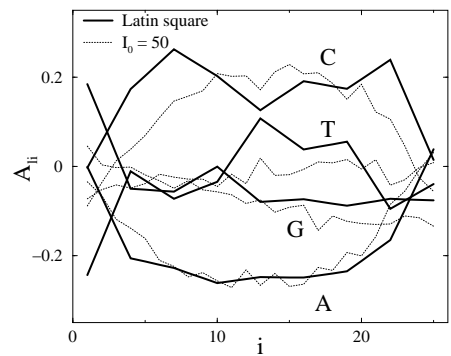}
\caption{Fit of Spike-in data of the HGU95A microarray using
eq.~(\ref{eq:naef}). Here, we subtract from the intensity the known
background intensity at zero concentration.}
\label{FIG03}
\end{figure}

The fact that the position-dependent affinities are lower for G than
for C and for A than for T is consistent with the hybridization
data in solution, as pointed out in Ref. \cite{carl05c_tmp}. This
apparent ``asymmetry" is due to the asymmetry between DNA strands of
the surface-bound probes and the RNA strands of the target molecules
in solution.

The fact that the effective affinities for G and T are close is quite
surprising, given the clear differences in binding free energies in
solution; we will argue below that this is due to hybridization between
RNA target molecules in solution.

\section{Effective affinities resulting from molecular based models}
\label{sec:affymodeldata} 

To obtain more insight into the relation between the hybridization
free energies of Table \ref{table_DG} and the effective affinities of
Refs. \cite{naef03,bind05} and which we analyzed in the previous section,
we extract effective affinities from a model which was recently proposed
by two of us \cite{carl06}.

This model is based on ideas from Held \etal \cite{held03}.  As it
uses as input the binding free energies between DNA and RNA strands
in solution reported in Table \ref{table_DG}, we will refer to it as
the molecular-based model.  Additionally, it incorporates the effect
of binding in solution of RNA to RNA in an approximate way, fitted to
the intensities measured on an Affymetrix chip.  The intensity $I_s$ of
sequence $s$ is assumed to be proportional to the fraction of hybridized
probes at the surface, described by a Langmuir model.  In detail, it is
given by \cite{carl06}
\be
I_s = I_0 + \frac{\alpha_s c_s Z_s}{1+\alpha_s c_s Z_s}
\ I_{\text{max}},
\label{fluorescence}
\ee
where $c_s$ is the total concentration of targets with sequence $s$ in
solution, $Z_s$ is the partition sum over states in which target $s$
is bound to the probe, and $\alpha_s$ is the fraction of targets in
solution which are free, and not hybridized in solution.

\begin{figure}[h]
\begin{center}
\includegraphics[width=2.8cm]{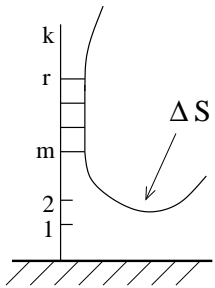}
\caption{Schematic picture of a partially hybridized configuration.
The total probe length is $k$ base pairs and we allow for $k < 25$, as
the photolitographic process used by Affymetrix produces probes which
are polydisperse. The target here is bound partially from bases $m$ to
bases $r$. We include in the calculation the entropy loss $\Delta S(m)$
due to the proximity of the target tail and the surface.}
\label{FIG04}
\end{center}
\end{figure}

In the model of Ref. \cite{carl06} 
\be
Z_s = \exp(-\beta \Delta G_s),
\label{part_func_HC}
\ee
where $\beta=1/(RT)$ is the
inverse temperature, and $\Delta G_s$ is the total binding free energy
for a perfectly formed helix of 25 base pairs between the RNA target
and DNA probe. This binding free energy is described by
\be
\Delta G_s = \sum_{ill'} S_{l,i} S_{l',i+1} \Delta G (l,l') + \Delta \Ginit.
\label{stacking}
\ee
As before, $S_{l,i}$ is a boolean variable equal to 1 if the probe
sequence has letter $l$ at position $i$ and 0 otherwise.  Thus, the sum
in eq.~(\ref{stacking}) runs over all 24 stacking parameters $\Delta
G(l,l')$, which depend on the identity of two neighboring nucleotides $l$
and $l'$ in the surface-bound DNA strand. $\Delta \Ginit$ represents a
helix initiation cost \cite{bloo00}. For the stacking parameters the model
uses RNA/DNA free energies given in Table \ref{table_DG}, as obtained
from experiments in solution~\cite{sugi95_sh}.  Note that, differently
from the approach of Refs. \cite{naef03} and \cite{bind05}, the free
energies used here are position-independent.  In Ref. \cite{carl06},
the inverse temperature $\beta$ in eq~(\ref{part_func_HC}) is taken as
a fitting parameter.

We stress that in Ref. \cite{carl06} the hybridization free energy
$\Delta G = \Delta H - T \Delta S$ was taken at $T=37^\circ$C, while an
Affymetrix hybridization experiment is performed at $T=45^\circ$C, which
is the value we consider here (see Table \ref{table_DG}).  Although the 
temperature differs by only $8^\circ$C, the $\Delta G$'s on average differ
by about $20\%$, since $\Delta H$ and $T \Delta S$ are rather close. We
took the sequences of the Latin square set (25 nucleotides of length)
and generated $\Delta G$ of each sequence at both temperatures. A plot
of $\Delta G_{37}$ vs. $\Delta G_{45}$ shows that the values are narrowly
distributed along a straight line.  This implies that a difference between
the two choices of parameters can be reabsorbed in a rescaling of $\beta$
in eq.~(\ref{part_func_HC}).

Of practical interest is the total concentration $c_s$ of targets
with sequence $s$. Due to hybridization of single-stranded RNA in
the solution, the concentration of free targets, which can bind to the
probes, is lower than the total concentration of targets in solution.
In the model of Ref. \cite{carl06}, this is taken into account
by reducing the total concentration $c_s$ in solution by a factor of
$\alpha_s$ given by
\be
\alpha_s = \frac 1 {1 + c_0 \exp{(\beta' \Delta G_R)}},
\label{alpha}
\ee
where $\beta'$ and $c_0$ are fitting parameters and $\Delta G_R$ is
the (sequence dependent) RNA/RNA binding free energy for duplex formation
in solution, taken from Ref.~\cite{bloo00}.  Note that also $\alpha_s$
is highly sequence-dependent: CG-rich targets will have high affinity
to the complementary surface bound probes, but will also have a strong
tendency to hybridize in solution.  It has been shown that a unique
choice (i.e. probe-independent) of the parameters $I_{\rm max}$, $\beta$,
$\beta'$ and $c_0$ fits the experimental data well \cite{carl06}.

\begin{figure}[h]
\includegraphics[width=8.3cm]{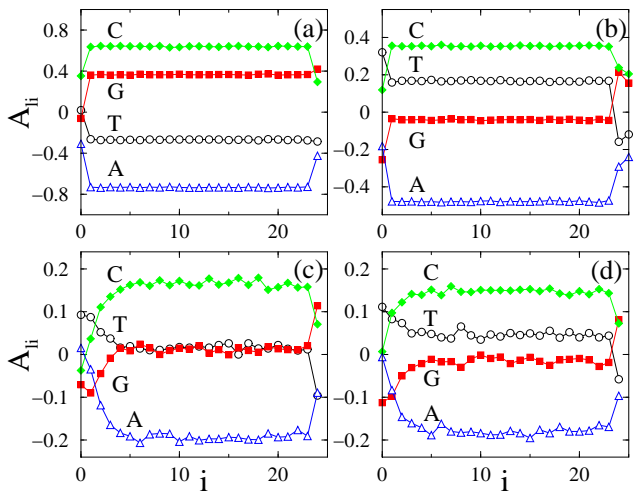}
\caption{Effective affinities, obtained with the molecular-based model,
versus position in the probe, for the four different nucleotides.
In panel (a) only the binding energy is taken into account; the effective
temperature $T$ is 800 K. In \fig (b), the hybridization in
solution is also taken into account, as in the molecular-based model of
Ref. \cite{carl06}; the resulting effective temperature becomes
$T=570$ K. The effect of using the ``zipper'' and the probe
length distribution is shown in \fig~(c), resulting in an effective
temperature of $T=525$ K. In \fig~(d) all effects mentioned in the
text are taken into account and the effective temperature is reduced to
$T=494$ K.}
\label{FIG05}
\end{figure}

There are many similarities, and also some discrepancies,
between the intensities $I_s$ in the Naef and Magnasco (NM)
approach eq.~(\ref{eq:naef_bg}) and in the molecular-based model
eq.~(\ref{fluorescence}). The binding free energy in the NM approach is
captured in the summation on the right-hand-side of eq.~(\ref{eq:naef}),
which is very similar to the summation in eq.~(\ref{stacking}) in the
molecular-based model.  NM uses a summation over single base pairs with
position-dependent affinities, while the molecular-based model uses (in
eq.~({\ref{stacking})) a summation over pairs of base pairs (allowing for
stacking energies), with a position-independent strength.  As we already
mentioned, NM does not feature saturation, while the molecular-based
model does through the denominator in eq.~(\ref{fluorescence}).  Finally,
the clear position-dependence in the effective affinities obtained
with the NM approach is not included in the molecular-based model of
Ref.~\cite{carl06}.

\subsection{Extending the molecular-based model}
\label{subsec:extension}

In this work, we introduce several extensions to the latter model. These
extensions will cause position-dependence in the effective affinities,
without ad-hoc modifications to the stacking free energy parameters.
Most of these extensions are related to the fact that both target and
probe are polydisperse in length, and that the duplex can fluctuate
and partially unzip.  We will first explain these extensions, and then
discuss their effect later.
\begin{itemize}
\item {\em Unzipping.} Besides the configuration in which the target
is bound to the probe over its full length, other configurations
occur in which the target covers only part of the probe.  This is
taken into account by a ''Zipper''-model. As a result, the partition
sum $Z_s$ does not only contain a single term $\exp(-\beta \Delta
G_s)$, but is a summation over many terms, each of which given by
eq.~(\ref{stacking}), but in which the index $i$ runs from the first
bound pair $m$ to the last bound pair $r\ge m$.  This idea is visualized
in \fig~\ref{FIG04}.

\item {\em Probe length dispersity.} During the production process of
the Affymetrix chips, the probability $p_g$ that the probe grows by an
extra nucleotide is only around $p_g \approx 90\%$ \cite{form98}. This
means that the fraction of probes which reach the final full length of
25 nucleotides is $P(25)=(p_g)^{25}$.  The fraction of incomplete probes
reaching a length $l<25$ equals $P(i)= (p_g)^l (1-p_g)$.  We have included
the effect of probe length dispersity by including these probabilities in
the calculation. The intensity is therefore equal to $I= \sum_{l=1}^{25}
P(l) I_l$, where $I_l$ is the Langmuir isotherm corresponding to a probe
of length $l$.

\item {\em Non-specific binding.} Even in Affymetrix experiments where no
perfect matching targets are present, the intensity does not fall well
below 0.5\% of the maximal intensity. We attribute this to non-specific
binding to the probes. To account for the non-specific binding, 
we include in our model a constant sequence-independent background intensity
$I_0$.

\item {\em Tail repulsion.} The RNA-target molecules often
extend beyond the 25 base pairs of the probe; the average target
length is 50 base pairs.  The tail of the target which sticks
out from the base of the probe is hindered significantly by the
surface (see \fig~\ref{FIG04}).  This causes an
entropic repulsion between the target and the surface, lowering
the intensity.  The mathematics of this effect is presented in
Appendix~\ref{sec:appentropy}. This effect is not sequence-dependent
and the parameters $Z_s$ in eq.~(\ref{fluorescence}) can therefore 
be multiplied by a constant factor $Z_{S_{\text{tail}}}$, given in
eq.~(\ref{Ztailentropy}).

\item {\em Fluorescent labels.} 
Due to the fact that in the experiments only the U and C nucleotides can
have a label, the fluorescence intensity will scale linearly with the
number of U and C nucleotides, which obviously depends on the sequence.
We therefore multiplied each Langmuir isotherm by for a factor $2X_{UC}$,
in which $X_{UC}$ is the fraction of U and C in the target sequence. We
assumed that the target is simply composed of 25-mers.
\end{itemize}

\subsection{Results of the model calculations}

We generated 100,000 different random sequences of 25 nucleotides each.
For each sequence $s=1\dots 10^5$, we also selected a concentration
$c_s$, with a minimal value of $c_{min}=1$ picomolar and $c_{max}=$
1 nanomolar (the typical range of target concentrations in Affymetrix
arrays); the logarithm of these concentrations $\log(c_s)$ is drawn
from a uniform distribution $[\log(c_{min}),\log(c_{max})]$.  For each
sequence $s$, the intensity $I_s$ is calculated using the molecular-based
model, eq.~(\ref{fluorescence}), with the extension just described.
The parameters entering this equation are the stacking free energies given
in Table~\ref{table_DG}, as well as the parameter $\alpha_s$ reflecting
the reduction of the total concentration of free targets in solution;
this latter (sequence-dependent) parameter uses the RNA/RNA binding free
energies for duplex formation in solution, taken from Ref.~\cite{bloo00}.
The modifications in the molecular-based model as compared to the model
in Ref.\cite{carl06}, as well as the different choice of free energy
parameters ($\Delta G_{45}$ vs $\Delta G_{37}$) require a refitting of
the effective inverse temperature $\beta'$ and a concentration $c_0$,
which yielded $\beta'=0.6 (\mbox{kcal/mole})^{-1}$ and $c_0=e^{\epsilon
\beta'}$, with $\epsilon=42$ kcal/mole.  The fitting procedure for these
two parameters follows the procedure of \cite{carl06}.

In the experimental Affymetrix data set, the average intensity is around
3\% of the maximal intensity.  In all our simulations, we adjusted
the temperature to reproduce this average intensity.  The resulting
temperatures range from 494 K to 550 K. There is still a gap between the
experimental temperature of 318 K, but including the effects mentioned
above has significantly decreased this gap in the original molecular-based
model, where the effective temperature was 700 K \cite{carl06};
in turn the latter model had already a much more realistic effective
temperature than the Held model where the effective temperature exceeded
2000 K \cite{held03}.  
To obtain the effective affinities $A_{li}$ associated to the
molecular-based model, we minimize the difference between the
intensity 
$I_s$ as predicted by the molecular model in eq.~(\ref{fluorescence}) and the 
intensity $I_e$ resulting from the effective affinities and given by
\be
\ln(I_e)=\sum_{li} S_{li} A_{li} -\ln(c_s),
\ee
in analogy to eq.~(\ref{eq:naef}). More precisely, the effective affinities
$A_{li}$ result from a minimization of the sum over all 100,000 sequences
of the the squared difference between the logarithm of the intensity $I_s$
and the logarithm of the intensity $I_e$ resulting from the effective
affinities.

The first data set comprises a simple two-state model, in which a target
is either free in solution, or fully bound to a probe.  Hybridization in
solution is not taken into account, i.e., $\alpha_s=1$.  The results are
shown in \fig~\ref{FIG05}(a).  The effective affinities do not depend on
the position, apart from the two edge nucleotides which enter in only
one pair of neighboring base pairs.  (See eq.~(\ref{stacking})). Note
that the affinities increase with the ordering $A<T<G<C$, as expected
from the values of the free energies of Table \ref{table_DG}.

Next, the hybridization in solution is taken into account by using two
extra parameters $\beta '$ and $c_0$ which have the values of $\beta'=0.6
(\mbox{kcal/mole})^{-1}$ and $c_0=e^{\epsilon \beta'}$ with $\epsilon=42$
kcal/mole, respectively. Still the effective affinities are not position
dependent, see \fig~\ref{FIG05}(b). However, the order of the curves
has changed: $A<G<T<C$.

Fig.~\ref{FIG05}(c) shows the effective affinities when
polydispersity of the probe length distribution and the effect that a
duplex can zip open has been taking into account. These two effects lead
to position-dependent effective affinities. The effect on the side of the
microarray surface is larger than that on the solution side. Furthermore,
the effective affinities of G en T have become more alike.

The last panel of \fig~\ref{FIG05} shows the effective affinities when
also the effect of noise, entropy of the tails, and the fact that only
U and C carry fluorescent labels are taken into account. The biggest
effect is that the effective temperature is lowered.  Furthermore,
the sequence has become $A<G\approx T<C$, in agreement with the order
of effective affinities observed in experiment (see \fig.~\ref{FIG02}).
Note also that the scale of amplitudes of the effective affinities ranges
from about $-0.2$ to $0.2$ (see Fig.~\ref{FIG05}(c-d)). This is fully
consistent with the values obtained in Section \ref{sec:affyexpdata}.

\section{Conclusions}
\label{sec:conclusion}

In this paper we have analyzed the relation between the effective
affinities as originally introduced by Naef and Magnasco \cite{naef03}
and those obtained from a molecular-based model \cite{carl06}, which
uses hybridization free energies in solution. We show that these two
models yield very similar effective affinities.  This implies that free
energies in solution are adequate parameters to describe hybridization
in Affymetrix microarrays, at least if an effective temperature is used.

Firstly, the fact that the effective affinity for G is lower than
that for C and that the affinity for A is lower than that for T is
consistent with hybridization data in solution, as pointed out in
Refs. \cite{bind05,carl05c_tmp}.  Here, we have shown the role of
target-target hybridization in solution, which in the molecular-based
approach \cite{carl06} is described by a parameter $\alpha$
(see eq.~(\ref{alpha})).  The effect of $\alpha$ is of diminishing the
differences in the effective affinities between different nucleotides
so to make the effective affinities for G and T almost ``degenerate"
(see Fig. \ref{FIG05}). This is consistent with the data of Naef and
Magnasco \cite{naef03}, Binder and Preibisch \cite{bind05} and our results
of Sec. \ref{sec:affyexpdata}.  The basic physics behind this effect is
quite clear. The small difference between the effective affinities for
G and T, in spite of the large difference in binding free energies in
solution between these two nucleotides, is caused by the fact that G-rich
sequences tend to hybridize strongly in solution, thereby diminishing
their concentration available for binding to the probes.

We note that the calculation of the previous section yield effective
affinities which are position-dependent, mostly caused by the ability
of the probe-target complex to partly open up at the ends. To a lesser
extent, also the target-surface repulsion and the polydispersity of the
probes play a role.  The profiles of the effective affinities calculated
in Sec. \ref{sec:affyexpdata} are somewhat smoother than those deduced
from the molecular-based model. This difference is however small. The
most important aspect of our analysis is however that the molecular-based
model 1) reproduces the degeneracy between the affinities of G and T
2) yield amplitudes for the affinities quantitatively close to those
calculated in Sec. \ref{sec:affyexpdata}.

We finally comment on other possible ways of linking effective affinities
to hybridization free energies obtained from melting experiments in
solution. A recent study Ref. \cite{bind04} attributed the differences
between the two quantities to the effect of biotin molecules on the
binding. This is an alternative point of view compared to our approach
which emphasizes instead the effect of hybridization in solution between
partially complementary single stranded RNA molecules. In this respect it
would be interesting if measurements of melting temperatures experiments
of biotinilated RNA and DNA duplexes in solution similar to that of
Ref. \cite{sugi95_sh} could be performed. These experiments would allow
to  quantify the effect of biotin on binding. To our knowledge such
experiments have not yet been performed.  

\acknowledgements

We acknowledge financial support from the Van Gogh Programme d'Actions
Int\'egr\'ees (PAI) 08505PB of the French Ministry of Foreign Affairs
and NWO grant 62403735.

\appendix 

\begin{figure}[h]
\begin{center}
\includegraphics[width=4.0cm]{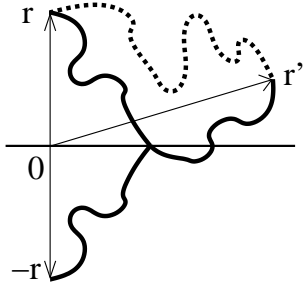}
\caption{The fraction of paths originating in $\vec{r} = (0,0,z)$ and
never crossing the plane $z=0$ can be found with the method of images:
the number of paths crossing the plane and ending in $\vec{r}\ ' $ is 
equal to the total number of paths starting from $-\vec{r}$ and
ending in $\vec{r}\ '$.}
\label{FIG06}
\end{center}
\end{figure}

\section{Entropic repulsion between substrate and target tail}
\label{sec:appentropy}

We model the single-stranded DNA segment as a freely jointed chain with
Kuhn length $b$. The probability distribution that a segment of $N$
Kuhn steps extends to a distance $\vec{r}$ from its origin is given
by a Gaussian distribution:
\be
\Gamma (\vec{r},N) = \left(\frac{3}{2 \pi N b^2}\right)^{3/2} 
\ e^{-3 r^2/2 N b^2}.
\ee
To determine the number of polymers starting from a height $z$ above
the surface and not crossing the wall, we use the method of mirror
images.  Using the same configuration as in \fig~\ref{FIG06}:
the fraction of walks of length $N$ originating from $\vec{r} = (0,0,z)$
and terminating at $\vec{r}\ ' = (0,0,z')$ is equal to
$\Gamma (\vec{r}\ ' - \vec{r},N)$. A part of these cross the wall.
This fraction is equal to $\Gamma (\vec{r}\ ' + \vec{r},N)$, i.e. the
number of walks originating in $-\vec{r}$ and terminating in $\vec{r}\ '$.
Therefore the fraction of walks of total length $N$ starting in
$\vec{r}$ and terminating in $\vec{r}\ '$ and which do not cross the
wall is given by the difference:
\beqn 
e^{\Delta S_N[\vec{r}]/R} &=& 
\int_{z'>0}  d \vec{r}\ ' 
\left[
\Gamma (\vec{r}\ ' - \vec{r},N) - \Gamma (\vec{r}\ ' + \vec{r},N)
\right]
\nonumber\\ 
&=& {\rm Erf}\left( \frac z b \sqrt{\frac{3}{2N}}\right),
\eeqn 
where Erf(x) denotes the error function defined as
\be
{\rm Erf} (x) = \frac 2 {\sqrt{\pi}} \int_0^x e^{-t^2} \ dt.
\ee
We recall that the Kuhn length is related to the persistence length as
$b = 2 l_p$ and that for single stranded DNA $l_p \approx 5$ bp.

We sum next over all possible tail lengths. Before hybridization the
target molecules are fragmented at random locations, with an average
fragment length of about $50$ bp. We find thus:
\be
e^{\Delta S(m)/R} = (1-\gamma) \sum_{N=0}^\infty \gamma^N 
{\rm Erf} \left[ \frac{m+m_0}{10} \sqrt{\frac{3}{2N}}\right],
\label{Ztailentropy}
\ee
in which $\gamma=49/50$ is the probability for chain continuation, and
$m_0$ is the ratio of the spacer distance and the length of a single
base pair.


\end{document}